\documentclass{article}
\usepackage{graphicx}
\usepackage{authblk}
\usepackage{subcaption}
\usepackage{xspace}

\usepackage[backend=biber,style=apa,sorting=nyt]{biblatex}
\usepackage[strings]{underscore}
\newcommand{\datasetname}{\textsc{FOSSIL}\xspace}

\title{Digging Up Citations: \datasetname, a Dataset and Workflow for Reference Extraction in Law and the Humanities \\ \large (Extended Abstract)}
\author[1]{Luca Foppiano\textsuperscript{*}}
\author[2]{Christian Boulanger\textsuperscript{*}}
\affil[1]{ScienciaLAB, Portugal}
\affil[2]{Max Planck Institute for Legal History and Legal Theory, Germany}
\date{January 2026}

\begin{document}

\maketitle
\begingroup
\renewcommand\thefootnote{\fnsymbol{footnote}}
\footnotetext[1]{These authors contributed equally to this work.}
\endgroup

\section{Introduction and motivation}

Citation extraction has been studied extensively~\parencite{colavizza2017annotated, colavizza2019citation, cioffi2022structured, backes2024comparing}, but existing models are designed for the structured end-of-document bibliographies typical of the natural sciences. In law and the humanities, scholars continue to cite references primarily in footnotes, where bibliographic data is interleaved with commentary, clarifications, and cross-references, producing highly heterogeneous content that is difficult to parse with standard pattern-based approaches~\parencite{boulanger2022extracting}.

Large Language Models show promise for such heterogeneous data~\parencite{simons2026large, sarin2025citation}, and even early zero-shot experiments produced strong results on footnote references~\parencite{textDavinci20222}. However, closed models depend on third-party infrastructure, are non-reproducible in the long term, and have been decommissioned without notice~\parencite{karkee2025comprehensive}; open-weight alternatives demand dedicated hardware. Meanwhile, LLMs are valuable in human-in-the-loop pipelines, where they accelerate semi-automatic annotation and enable larger datasets at lower cost. The limited availability of specialized gold-standard datasets for SSH literature hampers both the evaluation of LLM-based approaches and the training of classical supervised sequence labellers, making the construction of such resources a critical priority.

This work contributes: (i) \textbf{PDF-TEI Editor}, a production-ready web-based annotation tool; (ii) a documented \textbf{annotation workflow} integrating PDF-TEI Editor and Grobid~\parencite{grobid}, carried out by a team of seven; (iii) \datasetname (Footnote-based Open-access SSH Scientific Instance Labels), an openly licensed multilingual dataset of 96 annotated scholarly articles comprising over 7{,}600 references embedded in footnotes; and (iv) a \textbf{Grobid specialization} for law and humanities documents that serves as a proof-of-concept baseline.

\section{Background and related work}

Existing SSH-oriented datasets offer complementary but partial coverage. CEX~\parencite{cioffi2022data, pagnotta2024cex} covers 112 English articles but includes non-openly-licensed material. The EXparser Gold Standard~\parencite{hosseini2019excite} is restricted to 100 German social-science documents. LinkedBooks~\parencite{colavizza2017annotated} provides over 40{,}000 references from monographs and journals on Venetian history, but remains narrow in domain. \textcite{zhu2026benchmarking} benchmarked open-weight LLMs against Grobid and found that fine-tuned LLMs are more robust under distribution shift, particularly for non-English and footnote-heavy documents, recommending a routing strategy between Grobid and task-adapted LLMs.

\datasetname addresses remaining gaps by providing an openly licensed corpus that spans multiple SSH disciplines, languages, and citation styles, including the footnote-heavy conventions of law and humanities.

\section{Challenges of citation practices in law and the humanities}
\label{sec:challenges}

In natural-science publications, references form a dedicated section with standardised inline markers, while footnotes are reserved for commentary. In contrast, law and humanities scholarship relies heavily on footnotes to carry bibliographic references alongside commentary, and citation styles vary strongly across languages, disciplines, and journals. Because footnotes are page-scoped, the same reference frequently recurs in shortened form as a cross-reference to an earlier footnote.

We identified five recurring footnote content types: (1) comments or clarifications with no reference data; (2) biographical information, typically on the first page; (3) one or more bibliographic references that can be cleanly segmented; (4) cross-references (e.g., ``Ibid.'', ``See Miller, n.\,9, p.\,5'') redirecting to a previous footnote; and (5) mixed content combining all of the above in arbitrary order.

\section{Method}

Our approach is grounded in the Grobid architecture: a cascade of specialized machine-learning models extracting and parsing PDF articles into structured TEI-XML. \datasetname is aligned with this layered design so that annotations can be used independently at each stage or combined end-to-end.

\paragraph{PDF-TEI Editor.} We developed an open-source web application supporting iterative, collaborative TEI-XML annotation~\parencite{boulanger2024pdftei}. It features a synchronized dual-pane PDF/XML view, built-in TEI schema validation, role-based access control, and a plugin architecture that allows integration of arbitrary extraction backends (currently Grobid and LLaMore~\parencite{llamore}). Versioning with branching, merging, and change tracking supports the iterative workflow described below.

\paragraph{Annotation workflow.} Source articles were drawn from openly licensed journals listed in DOAJ and CrossRef, strictly excluding non-derivative licences. Documents were organised into ten batches and annotated in five phases: (1) \emph{pre-annotation} with the latest Grobid model; (2) \emph{correction} by a first annotator and verification by a second; (3) \emph{validation} by a senior reviewer; (4) \emph{feedback} integration into a living annotation guide, with GitHub issues used to track decisions; and (5) \emph{training and evaluation} of a new model. Seven annotators participated: two senior annotators (the authors), four students from the Digital Humanities Master Program at the University of Stuttgart, and one reviewer.

For segmentation, 194 pairwise comparisons on 96 documents yielded a mean Krippendorff's~$\alpha = 0.917$ (median $0.989$), indicating reliable agreement, with a clear learning curve across batches.

\paragraph{Dataset.} \datasetname comprises 96 articles from 38 journals in law, humanities, history, and social sciences, with publication dates from 1959 to 2025 and multiple languages (English, Portuguese, Italian, German). The corpus contains approximately 2{,}400 footnote blocks---about 500 discursive and 1{,}900 bibliographic---comprising over 7{,}600 individual references, averaging roughly 100 per document. Annotations are organised in three layers matching the Grobid cascade: article segmentation (structural zones such as \texttt{<front>}, \texttt{<body>}, \texttt{<listBibl>}, \texttt{<note>}), reference segmentation (splitting \texttt{<listBibl>} content into individual \texttt{<bibl>} units), and citation parsing (TEI-standard fields extended with a \texttt{<note>}-based encoding for cross-references). The dataset is released both in aggregated form and as separate files per layer.

\paragraph{Grobid specialization.} We implemented a specialized processing flow (``flavor'') that incrementally replaces models in the standard Grobid cascade---starting with article segmentation---while reusing the rest of the pipeline. This design follows a principle of maximum reusability and enables targeted evaluation of each replaced component.

\section{Evaluation}

We evaluate the trained article segmentation model in isolation and the full specialized pipeline end-to-end against the default Grobid configuration.

\paragraph{Article segmentation.} Using 5-fold cross-validation on 96 documents, the specialized segmentation model reaches an overall F1 of 81.08 (Precision 82.91, Recall 79.41). Performance is strong for \texttt{<body>} (86.85), \texttt{<headnote>} (80.89), and \texttt{<page>} (90.12), while \texttt{<listBibl>} (78.91), \texttt{<footnote>} (54.95), and \texttt{<front>} (34.54) are more difficult: these zones share physical page space and can only be distinguished by content, making classification inherently ambiguous.

\paragraph{End-to-end evaluation.}
\label{sec:end-to-end-evaluation}

To assess the impact of the specialized flow on real-world processing, we ran an end-to-end evaluation on four scholarly articles from law and the humanities, previously used to compare an LLM-based extraction library against standard Grobid~\parencite[see][p.~X]{boulanger2026potential}. Each document was processed through the full pipeline---from PDF to TEI-XML---under two configurations: default Grobid and the specialized ``footnotes-ref'' flavor. Extracted fields were compared against manually annotated gold-standard references using the LLaMore evaluation framework~\parencite{llamore}, which computes token-level F1 with fuzzy matching (Levenshtein threshold 0.8) and Hungarian matching for optimal reference alignment.

\begin{figure}[htbp]
  \centering
  \includegraphics[width=0.9\linewidth]{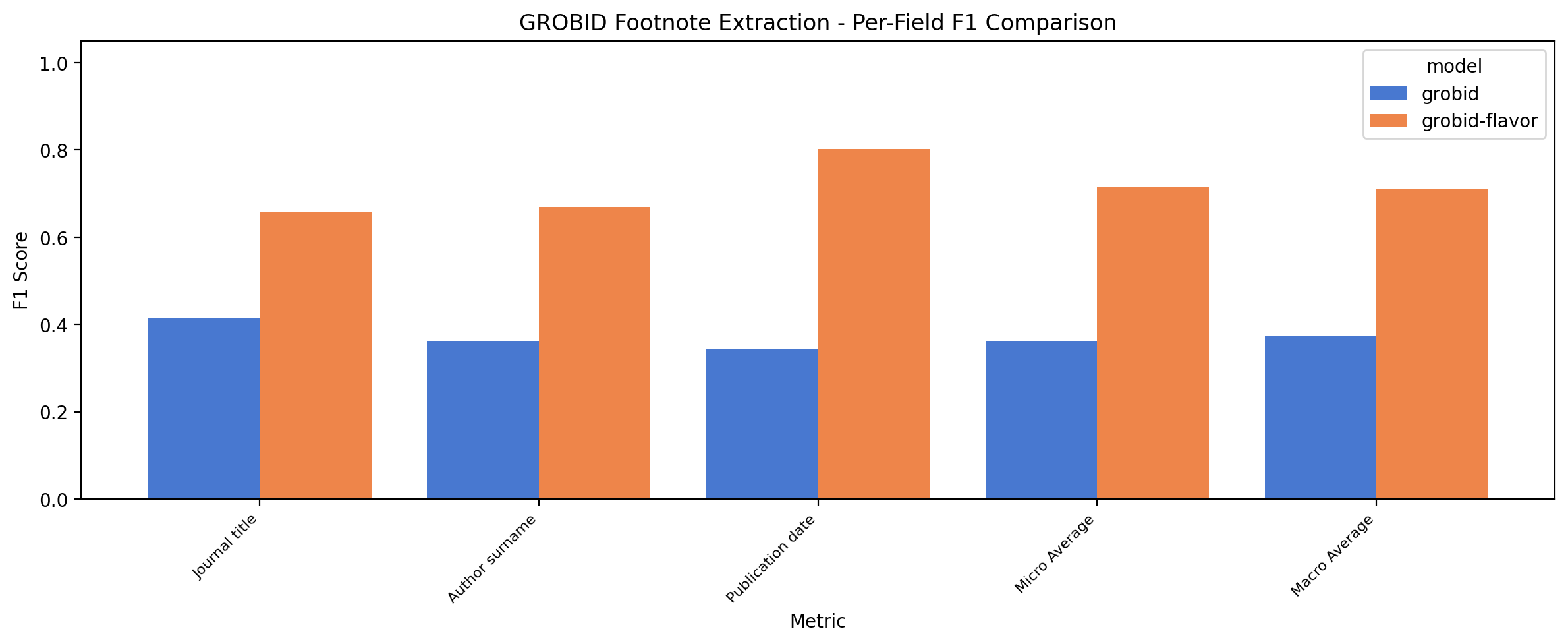}
  \caption{End-to-end evaluation on author surname, journal title, and publication date, comparing the standard Grobid pipeline and the specialized process that replaces only the segmentation model while reusing the full pipeline.}
  \label{fig:evaluation}
\end{figure}

Figure~\ref{fig:evaluation} reports per-field F1 for three key fields---journal title, author surname, and publication date---along with micro- and macro-averages.

The default Grobid model shows a highly asymmetric precision--recall profile: precision is high across all three fields (0.89--0.94), but recall is extremely low (0.21--0.27), producing F1 scores between 0.35 and 0.42. This reflects the fundamental mismatch between Grobid's default training assumptions and footnote-based citation styles: the few references it identifies are parsed correctly, but the vast majority of footnote-embedded citations are never recognised as bibliographic content.

The specialized process substantially corrects this imbalance. Recall improves across all fields---from 0.23 to 0.60 for author surnames, 0.27 to 0.73 for journal titles, and 0.21 to 0.71 for publication dates---while precision remains acceptable, yielding F1 scores of 0.67, 0.66, and 0.80 respectively. The micro-averaged F1 rises from approximately 0.36 to 0.72, a near-doubling of extraction quality.

The largest gain is on publication date (F1 0.35 $\rightarrow$ 0.80), reflecting the highly variable date conventions in legal and humanities scholarship (inline parenthetical dates, abbreviated years, dates interleaved with volume numbers). Author surname improves from 0.36 to 0.67, reflecting better handling of abbreviated forenames, particle names, and editorial attributions. Journal title improves from 0.42 to 0.66, consistent with better segmentation of citation strings lacking the explicit delimiters of scientific bibliography styles.

These results confirm that domain-specific specialisation of Grobid's sequence-labelling models substantially improves footnote citation parsing for law and humanities scholarship, while also showing that the primary failure mode---extremely low recall---cannot be addressed by post-processing alone. This motivates the dataset and workflow contributions of this work.

\section{Conclusion and future work}

We have presented \datasetname, together with an annotation tool, a documented workflow, and a Grobid specialization for footnote-based citation practices in law and the humanities. To our knowledge, \datasetname is the first openly licensed resource of this kind at this scale for legal scholarship, and the end-to-end evaluation confirms substantial gains from domain-specific specialisation (micro F1 from 0.36 to 0.72), while revealing significant headroom, particularly for cross-references and mixed-content footnotes.

Future work includes extending \datasetname to additional languages, legal traditions, and historical periods; tackling cross-reference resolution, which remains largely open; and exploring LoRA-based fine-tuning of LLMs~\parencite{zhu2026benchmarking} as well as end-to-end neural architectures that jointly optimise across article segmentation, reference segmentation, and citation parsing. \datasetname is intended as a stepping stone toward more expressive models capable of handling the full heterogeneity of SSH citation practices.

\section*{Acknowledgments}
We thank Dr.\ Anselm Küsters (Stuttgart University) and student assistants Mengjiao Li, Johanna Sandfort, Judith Schulze, and Marcia Sperber for their invaluable contribution to \datasetname.

\printbibliography

\end{document}